\documentclass[5p]{elsarticle}

 \usepackage{graphicx}

\usepackage{amssymb}

\usepackage[nodots]{numcompress}

 \usepackage{lineno}

\journal{Nuclear Physics B}

\begin{document}

\begin{frontmatter}



\title{Anisotropies in the  diffuse gamma-ray background measured by the Fermi-LAT}

 \author[label1]{A. Cuoco}
 \address[label1]{Stockholm University - Oskar Klein Center AlbaNova University Center, Fysikum, SE-10691, Stockholm, Sweden
}
 \author[label2]{T. Linden}
 \address[label2]{Department of Physics, University of California, Santa Cruz, 1156 High Street, Santa Cruz, CA, 95064}

 \author[label3]{M.N. Mazziotta}
 \address[label3]{Istituto Nazionale di Fisica Nucleare, Sezione di Bari, 70126 Bari, Italy}

 \author[label4]{J.M. Siegal-Gaskins}
 \address[label4]{Einstein Postdoctoral Fellow, California Institute of Technology 1200 E. California Blvd., Pasadena, CA 91125
}

 \author[label5]{Vincenzo Vitale\corref{cor1}}
 \address[label5]{Istituto Nazionale di Fisica Nucleare, Sezione di Tor Vergata, 00133 Roma, Italy}
\cortext[cor1]{E-mail: vincenzo.vitale@roma2.infn.it}

\author{for the Fermi-LAT Collab.}

 \author[label6]{E.Komatsu}
 \address[label6]{Texas Cosmology Center and Department of Astronomy, Univ. of Texas, Austin, Dept. of Astronomy, 2511 Speedway, Austin, TX 78712}



\address{}

\begin{abstract}

The small angular scale fluctuations of the (on large scale) isotropic gamma-ray background (IGRB) carry information about the presence of unresolved source classes. A guaranteed contribution to the IGRB is expected from the unresolved gamma-ray AGN while other extragalactic sources, Galactic gamma-ray source populations and dark matter Galactic and extragalactic structures (and sub-structures) are candidate contributors.

The IGRB was measured  with unprecedented precision by  the Large Area Telescope (LAT) on-board of the Fermi gamma-ray observatory,
and these data were used for  measuring the IGRB angular power spectrum (APS).
Detailed Monte Carlo simulations of Fermi-LAT all-sky observations were performed to provide a reference against which to compare the results obtained for the real data set. The Monte Carlo simulations are also a method for performing those detailed studies of the APS contributions of single source populations,
which are required in order to identify  the actual IGRB contributors.

We present preliminary results of an  anisotropy search in the IGRB.
At angular scales $<$2$^{\circ}$ (e.g. above multipole 155), angular power above the photon noise level is detected, at energies between 1 and 10~GeV in each energy bin,
with statistical significance between 7.2 and 4.1$\sigma$.
The energy not dependence of the fluctuation anisotropy is pointing to the presence of one or more unclustered source populations, while 
the energy dependence of the intensity  anisotropy is consistent with source populations having average photon index $\Gamma$ = −2.40$\pm$0.07.

\end{abstract}

\begin{keyword}


\end{keyword}

\end{frontmatter}


 \linenumbers

\section{IGRB Anisotropy}
In all-sky high-energy gamma-ray observations  an intense diffuse emission originating from the Milky Way is visible.
Above 30~MeV the large majority of this emission is produced
 by cosmic-ray (CR) ions interacting with the Galactic interstellar gas via neutral pion production and decay, and
inverse Compton (IC) scattering of interstellar radiation fields photons off CR electrons.
A fainter, almost isotropic on large angular scales, gamma-ray background (IGRB)
has also  been detected, first  by the SAS-2 mission \cite{fichtel}.
Later the IGRB energy spectrum was measured with good accuracy by the Energetic Gamma-Ray 
Experiment Telescope \cite{Sreekumar},\cite{strong} on-board
the Compton Observatory and recently with the LAT detector on-board  the Fermi
gamma-ray observatory \cite{abdoiso}. 

A considerable part of the IGRB has likely extragalactic
origin (EGB) and  carries information about the non-thermal phenomena in the universe.
The IGRB has a guaranted contribution from the known extragalactic gamma-ray sources.
AGN represent the largest source population above 30~MeV  detected by EGRET 
\cite{hart} and the LAT detector  
\cite{latcal1},\cite{latcal2}. Therefore, undetected AGN (those below the current
detection threshold) are the most likely candidates for the origin of the
bulk of the EGB emission. The estimates of the EGB fraction originating from AGN
vary from  20 and $\approx$100\% depending on the energy and the model (see \cite{abdoegb} and references therein).
Another main extragalactic candidate  is star-burst galaxies \cite{starburst}.
A fraction of the IGRB might also originate from the sum of Galactic nearby sources
such as millisecond pulsars (MSPs) \cite{MSPs}.

The IGRB anisotropy study provides us also with a method for the indirect search for 
dark matter (DM) with gamma rays, which is very complementary to the study of the 
main DM overdensities, such as the Galactic center \cite{vitale} and  
dwarf spheroidal galaxies \cite{dwarfs} and also complementary to  the search of  lines  \cite{lines} or other spectral features
from large areas of the sky.
In fact the sub-structures of the Milky Way dark matter halo and the 
dark matter halos in the local universe might produce an imprint in the IGRB.
Dark matter overdensities might shine in gamma rays both in the case of 
pair-annihilating DM particles (with the resulting gamma-ray flux proportional to 
the square of the dark matter density) and pseudo-stable decaying ones (with flux proportional to the density).
The largest substructures of the Galactic halo might be individually detectable while
the majority of them are likely to be under detection threshold but still be 
able to contribute to the IGRB \cite{ullio} and also induce small scale anisotropies \cite{JSG1}.

Information on the IGRB origins is carried by its small scale angular fluctuations. If a diffuse emission originates from
 unresolved source populations then small angular scale fluctuations will arise because of the variation of number density
 of the sources in different sky directions.
These fluctuations are a feature of the source distribution and will persist also in the limit of infinite statistics,
 contrary to Poisson fluctuations (photon noise)
which decreses with increasing event statistics.
It is therefore possible to discriminate between sources induced anisotropies and photon noise, 
if one is  provided with a sufficient data statistics.
One method for the study of the IGRB anisotropy is the APS measurement.
In theoretical studies of the IGRB anisotropy the following contributors were considered:
(a) blazars \cite{10},\cite{11},\cite{12}; 
(b) star-forming galaxies \cite{13}; 
(c)  Galactic MSPs  \cite{14};
(d)  annihilating or decaying   dark matter in Galactic sub-halos \cite{15},\cite{16},\cite{17};
(e) dark matter in extragalactic structures  \cite{9},\cite{10},\cite{12},\cite{17},\cite{18} and \cite{19},\cite{20}, \cite{21},\cite{22}.
In these studies it is also  shown that intrinsic clustering  of  many  populations candidates
has a sub-dominant effect on the angular power spectrum in multipole range between 100-500.

Here we report on a  search for anisotropies in the IGRB, performed with the data taken 
with the LAT detector on-board of the Fermi satellite.
\label{Context}

\section{The LAT and the Data Analysis}

The Fermi Large Area Telescope has a wide field of view (2.4 sr) and a large effective
 area ($\approx$8000 cm$^{2}$ for normally-incident photons above 1~GeV). 
LAT is a pair-conversion telescope with a modular structure made of 4$\times$4 $\emph{towers}$.
Each tower is composed by: (i) a precision silicon tracker (18 planes of Si-strip detectors
coupled with W conversion planes, with  a total of 1.5 X$_{0}$ for the normal incidence.
The Si tracker is made of a first thinner segment, called front, with a better 
angular resolution and a second thicker one, called back); 
(ii) a CsI homogeneous calorimeter (8.5 X$_{0}$ for the normal incidence). 
The pair-conversion telescope is  covered with an anti-coincidence detector that
 allows for rejection of charged particle events. Full details of the instrument, including
technical information about the detector, on-board and ground data processing, and
 mission-oriented support, are given in \cite{24}.

Data taken during the first 56.6Ms of observation ($\approx$22 months) were used.
The experimental data and simulations were analyzed with the LAT analysis software
 Science Tools version v9r15p4 with P6$_{-}$V3 LAT instrument response functions (IRFs).

The main analysis steps were:
\begin{itemize}
\item Data Preparation. By means of the gtselect tool:
(i) events of $\emph{diffuse}$ class were selected and with energy between 1 and 50~GeV; 
(ii) data with zenith angle exceeding 105$^{\circ}$ were rejected 
to reduce Earth gamma-ray albedo contamination;
(iii) events converted in the front and back tracker segments have also 
been divided in order to be analyzed  separately.
Periods in which with LAT was in the South Atlantic Anomaly,  not in survey mode or with rocking angle exceeding 52$^{\circ}$ 
were discarded with the tool gtmktime.
The  integrated livetime was calculated using 
the gtltcube tool (photon injection step size cos($\theta$) = 0.025,
 pixel size of 0.125$^{\circ}$ corresponding to a HEALPix \cite{23} map with Nside = 512 resolution)

\item Exposure. Exposure maps were calculated using the gtexpcube tool 
with the same pixel size as for gtltcube, for 42 logarithmic energy bins spanning 
1.04 to 50~GeV, in order to have a good knowledge of the energy dependence of the exposure.

The $\emph{event shuffling technique}$ is an alternative method for the exposure calculation 
that does not rely on the Monte Carlo based calculation of the exposure implemented in
the Science Tools. It was used for cross-checking  possible  exposure systematic errors.
With this method the arrival directions of pairs of detected events are swapped in the instrument coordinates, this produces anexposure map wih arbitrary normalization.
The same technique has also been used to search for anisotropy in the CR electrons arrival 
directions, measured with the LAT, in \cite{eanis}.

\item Intensity Maps. Counts maps were built with the selected data.
Both the photon counts and exposure maps were converted into HEALPix-format  maps
with Nside = 512, and   HEALPix gamma-ray   intensity maps in four energy bins were obtained.

\item Map Masking. Regions of the sky heavily contaminated by Galactic diffuse emission were excluded by masking Galactic
latitudes $|b| <$ 30$^{\circ}$, and masking sources in the Fermi 11-month catalog \cite{latcal1} within a 2$^{\circ}$ angular
radius. In this study we focused on multipoles $l\gtrsim$ 150 (corresponding to angular scales $<$ 2$^{\circ}$),
since lower multipoles (corresponding to correlations over larger angular scales) are likely more
contaminated by Galactic diffuse emission.

\item Intensity and  Fluctuation APS.
The angular power spectra were  calculated for the intensity map 
 I($\psi$), where  $\psi$ is the sky direction.
The angular power spectrum is given by the coefficients
C$_{l}$ = $<|a_{lm}|^{2}>$ with the a$_{lm}$ determined by expanding the map in spherical harmonics.
The intensity APS indicates the dimensionful amplitude of the anisotropy 
and can be compared with predictions for source classes whose collective
intensity is known or assumed.
A dimensionless fluctuation APS has also been calculated  by dividing the intensity 
APS spectrum C$_{l}$  of a map by the mean sky intensity (outside of the mask) squared $<$I$>^{2}$.
In the case of the shuffling technique only the intensity fluctuation  APS can be obtained.
The angular power spectra of the maps were calculated using the HEALPix package \cite{23}.
The measured APS were corrected for the power suppression due to the beam and
pixel window functions, and an approximate correction, valid at multipoles $l>$100, was applied
to account for the reduction in angular power due to masking. 
For each energy bin, the APS of the maps of front- and back-converting events were calculated separately and then
combined by weighted average.

\item Validation Studies. Careful checks of the procedures and the related parameters were
performed as well as tests for assessing the origin of the measured angular power.
The full analysis was applied to a simulated point source population, in order to compare the APS determined 
with the analysis chain and the one which could be analytically calculated. 
The dependence of the results on the  instrument response functions (IRFs), as also
on the masking have also been studied. For the latter in particular both the latitude cut with which the Galactic 
plane was excluded and the radius of the circle with which each know source was masked, were varied.
In order to further evaluate the effect of residual Galactic diffuse emission on the 
APS the analysis was repeated on the data after the subtraction of a model of the 
Galactic diffuse emission ($\emph{Galactic Foreground  Cleaning}$).
The details of these studies will be given in the final publication of the anisotropy study.
\end{itemize}

\begin{figure}
\centering
\includegraphics[width=8cm, height=5cm]{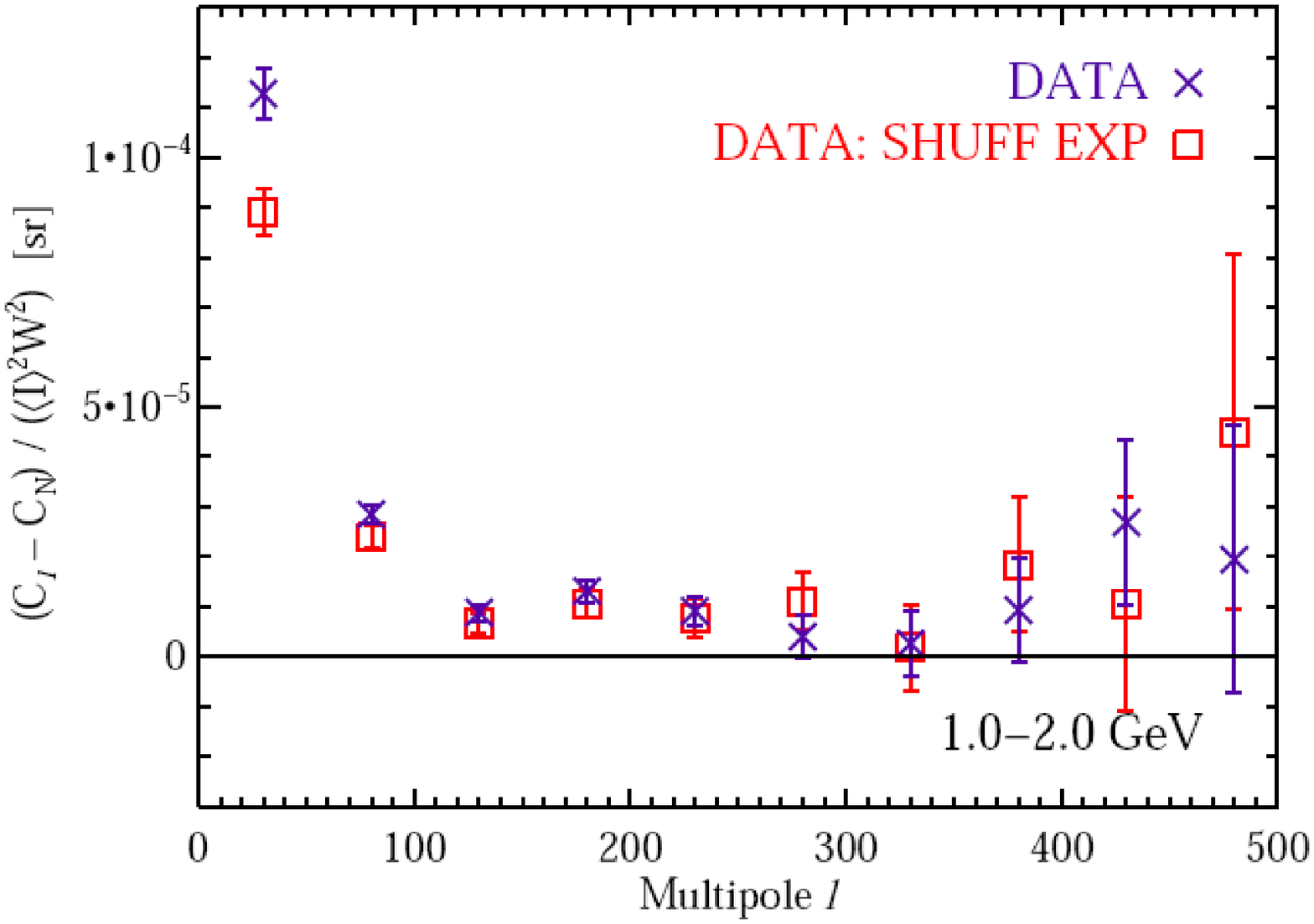}
\includegraphics[width=8cm, height=5cm]{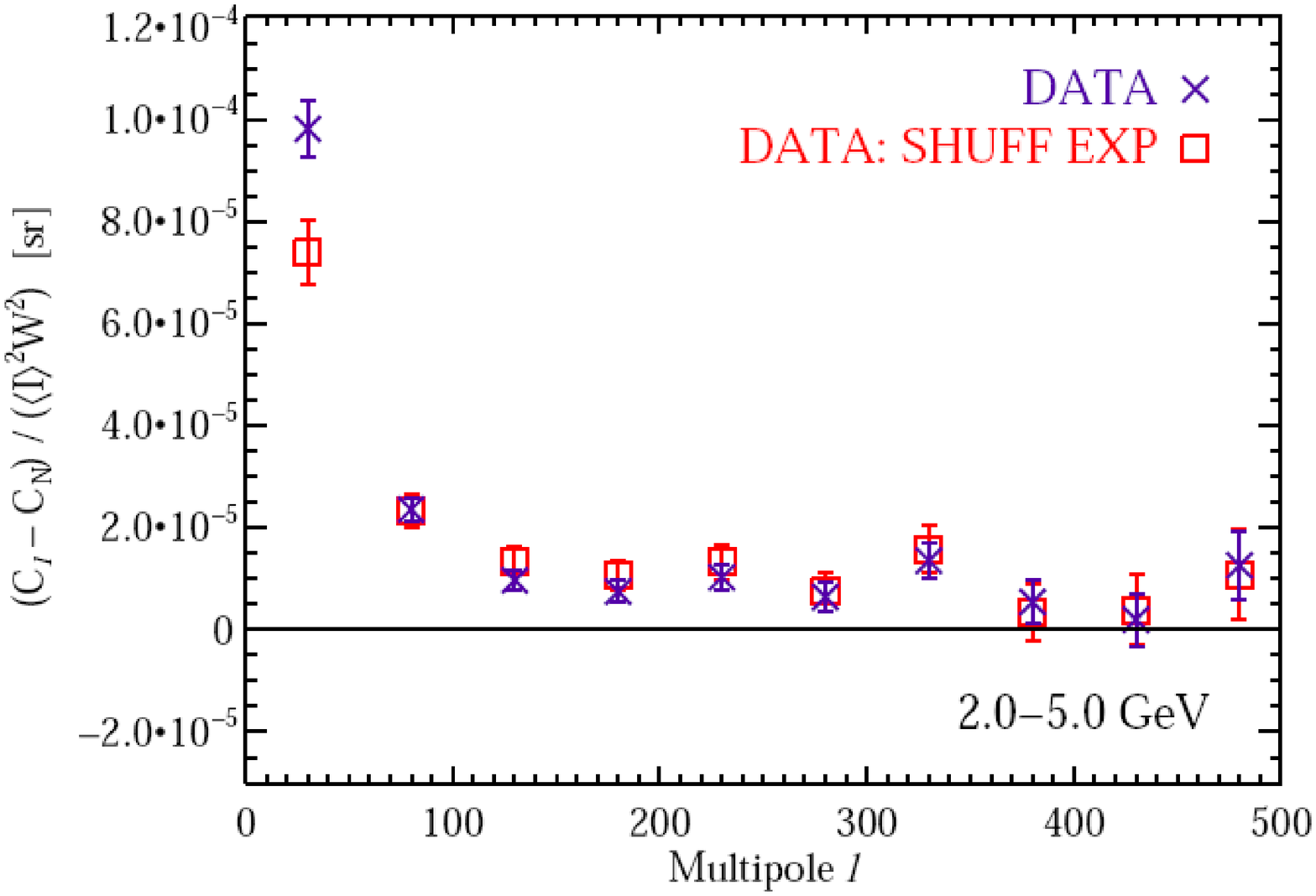}
\includegraphics[width=8cm, height=5cm]{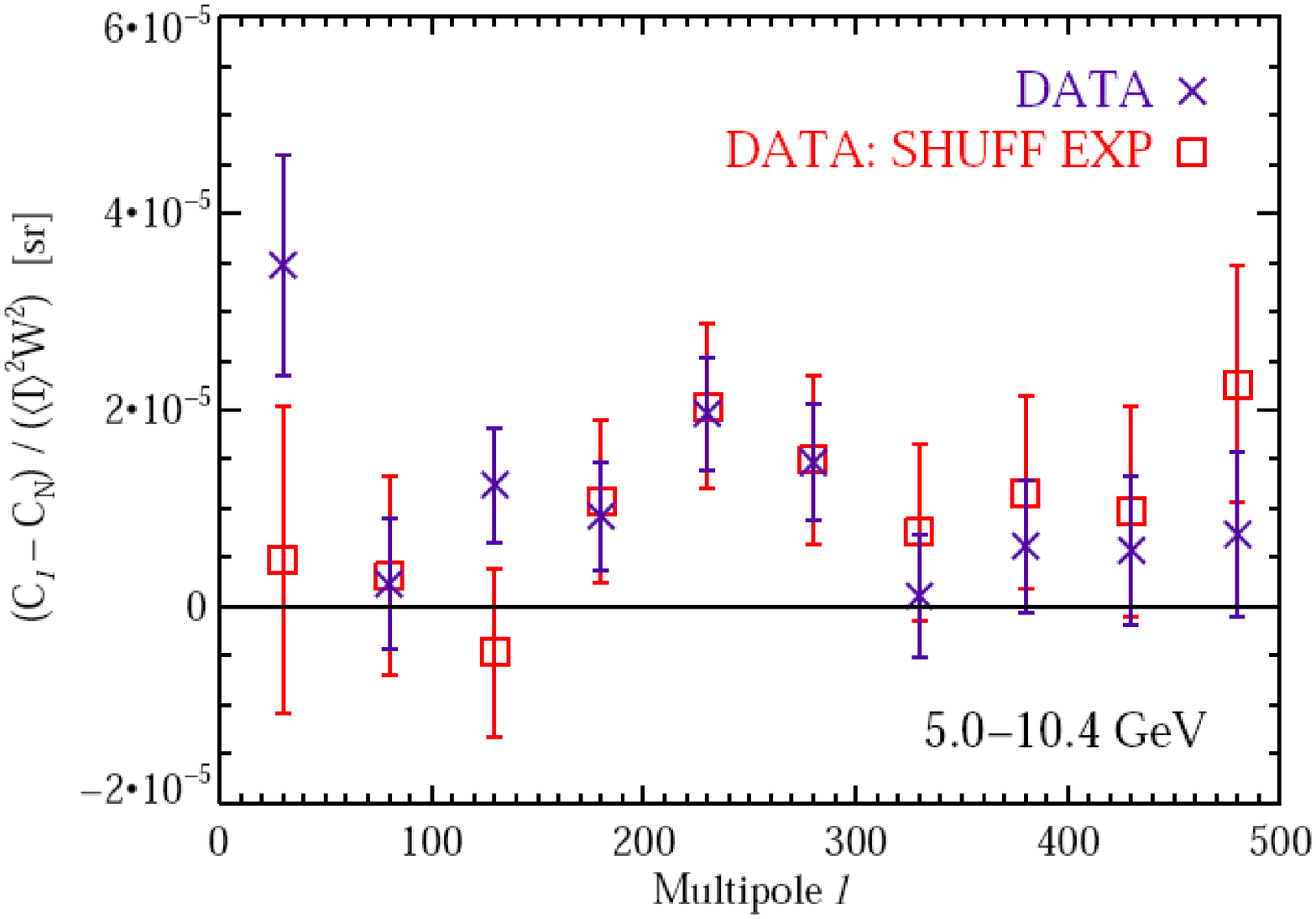}
\includegraphics[width=8cm, height=5cm]{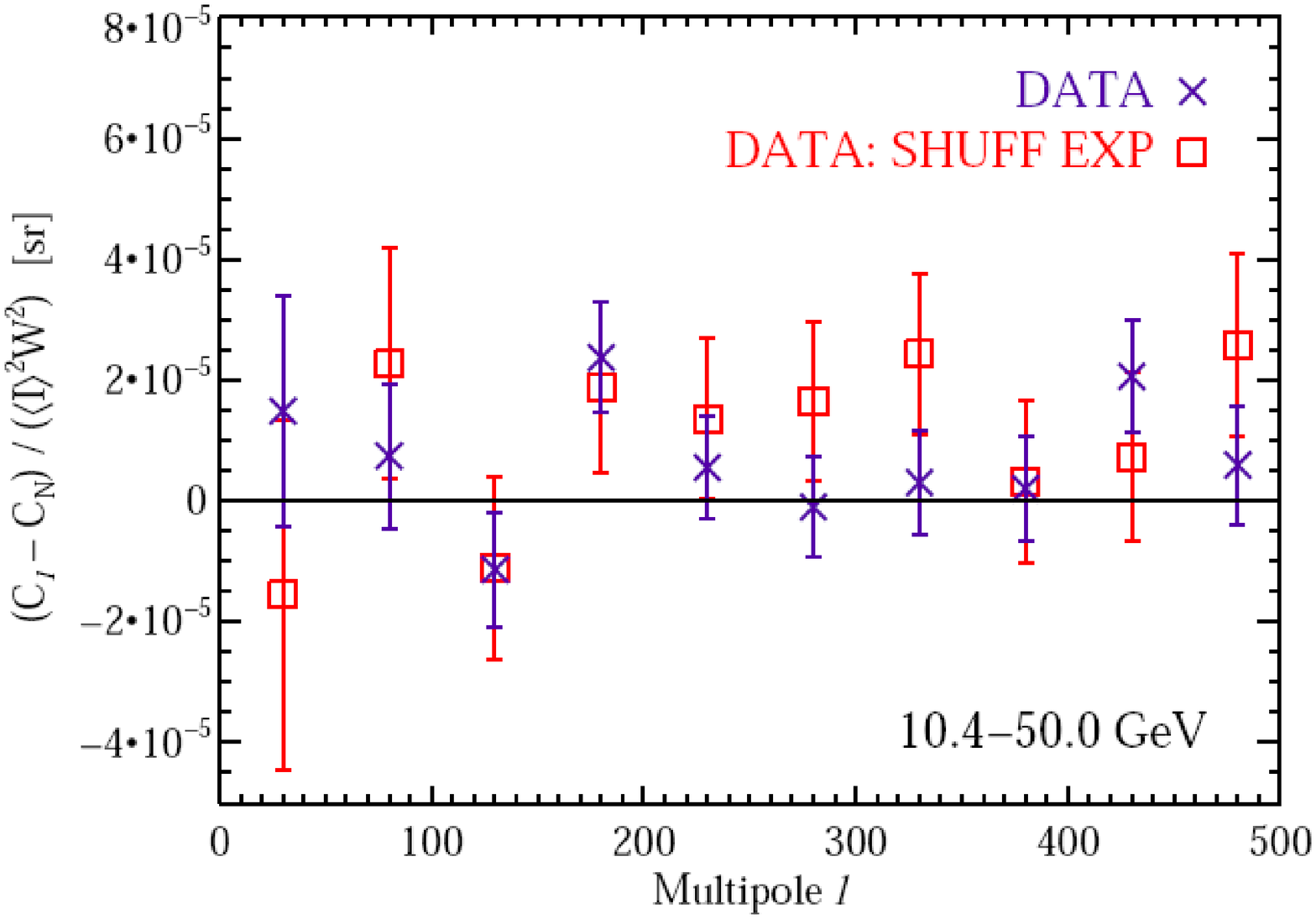}

\caption{PRELIMINARY. Fluctuation APS of the IGRB minus photon noise, in four energy bins.  For  isotropic emission this difference would be consistent with  zero. The large angular power at $l<$ 155  is likely originating from Galactic diffuse emission. For $l>$ 155 the measured power excess is approximately constant in multipole, suggesting that it originates from one or more unclustered source populations. The APS obtained with the event shuffling technique are also reported. The two different methods provide consistent results and show that possible inaccuracies in the exposure calculation have a negligible impact on these results.}
\label{fig:1}
\end{figure}
\begin{figure}
\centering
\includegraphics[width=8cm, height=5cm]{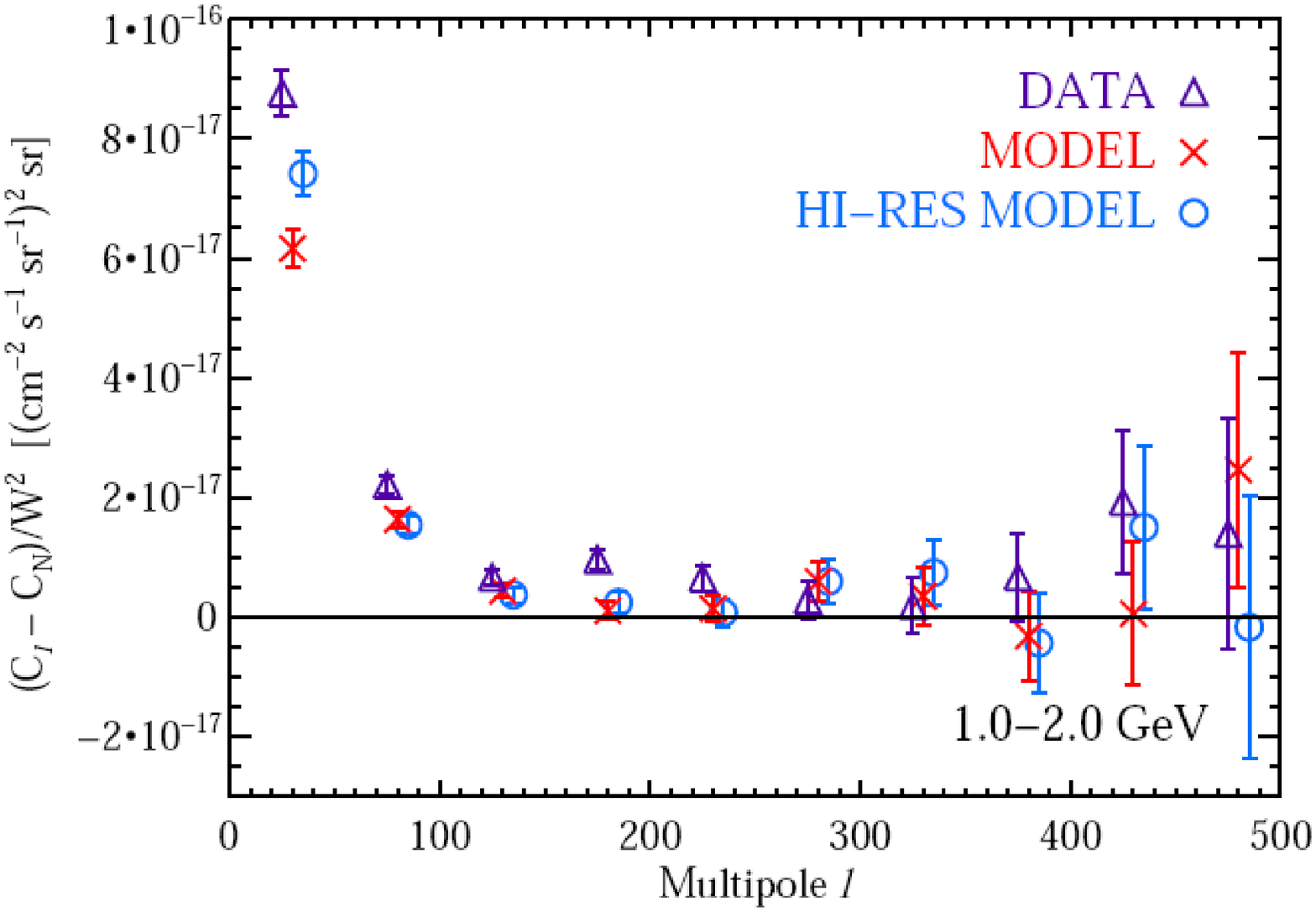}
\includegraphics[width=8cm, height=5cm]{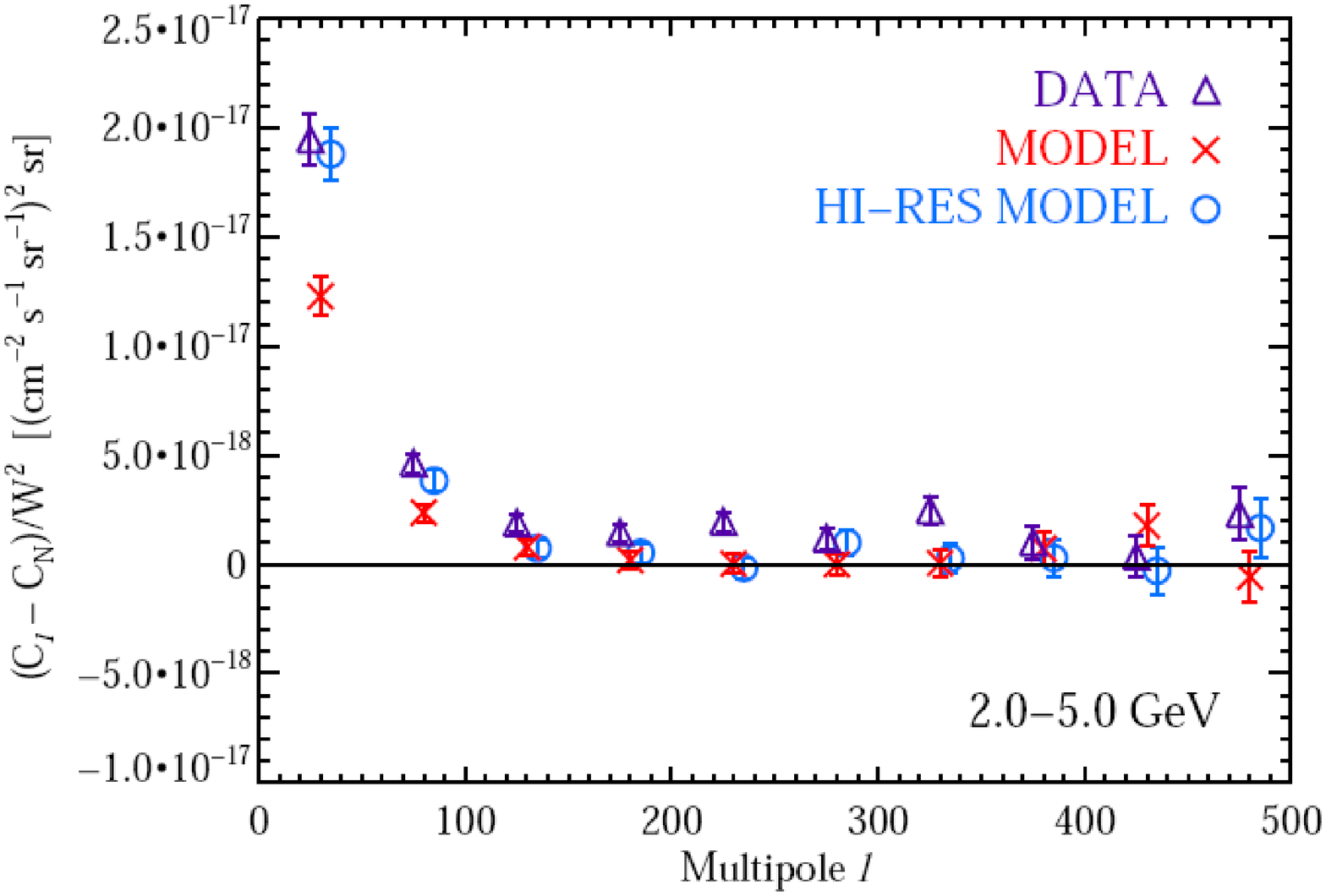}
\includegraphics[width=8cm, height=5cm]{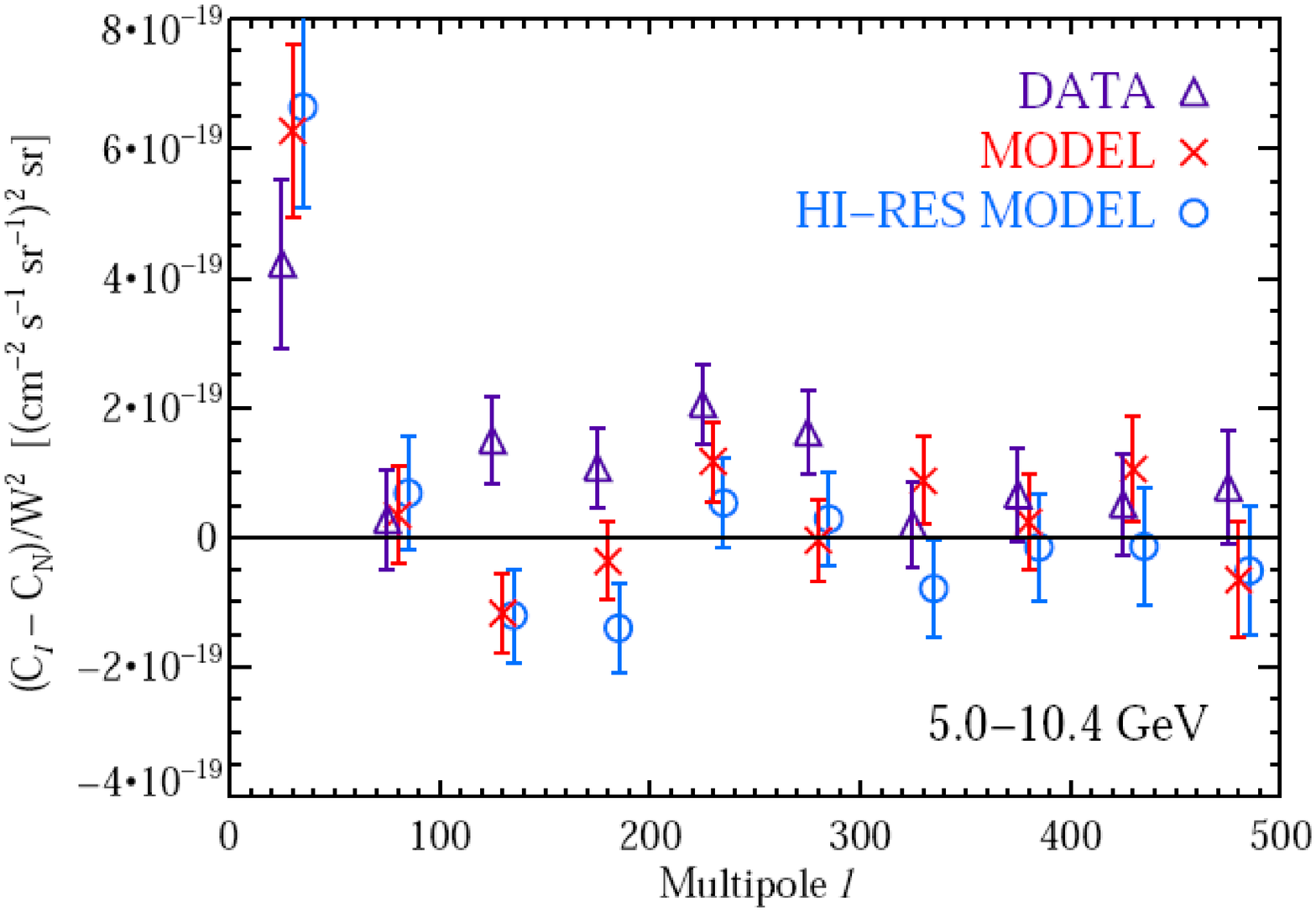}
\includegraphics[width=8cm, height=5cm]{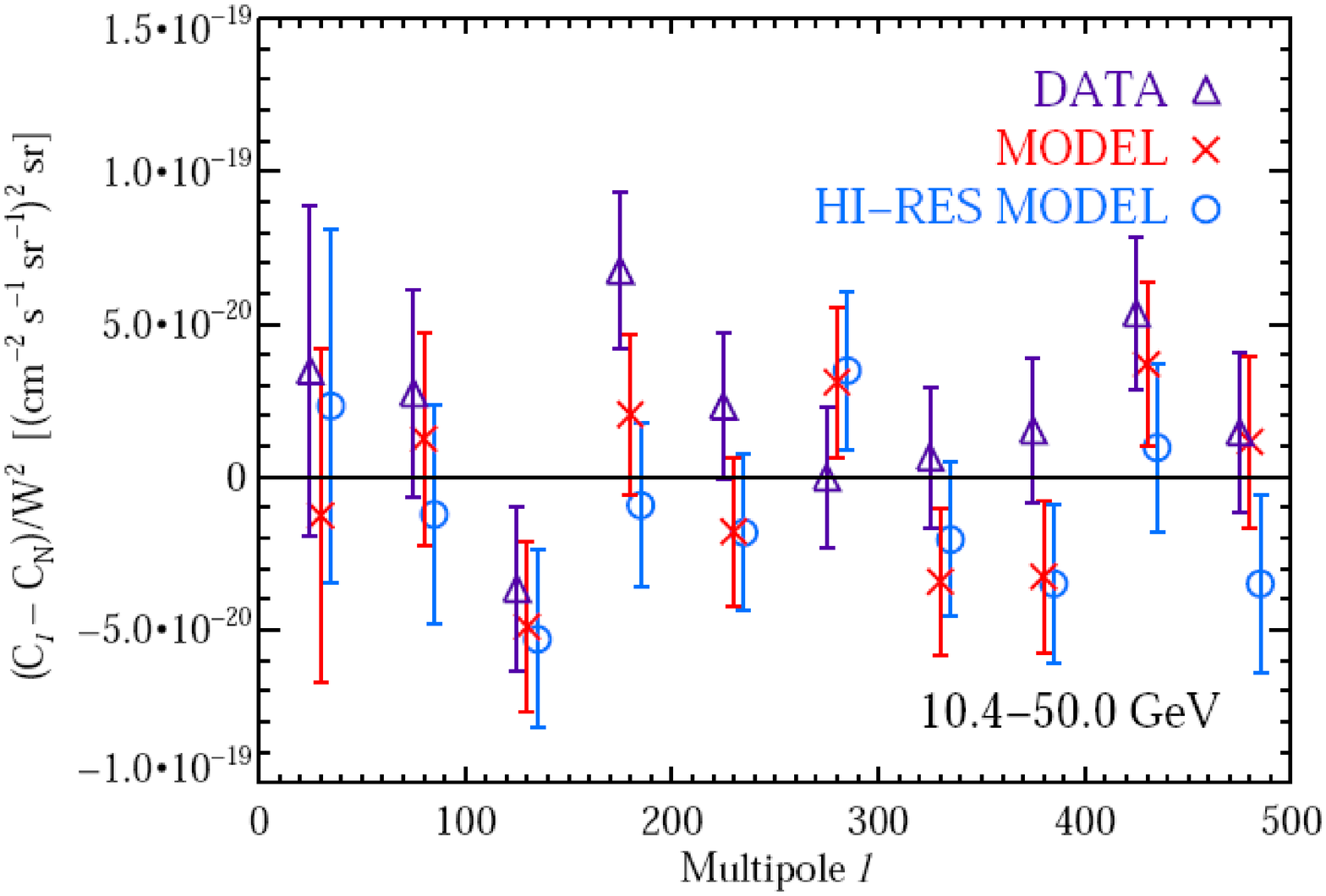}
\caption{PRELIMINARY. Intensity APS of the IGRB minus photon noise, in four energy bins. APS of the experimental data, simulated default model and high resolution models  are reported. The measured power above l $\approx$ 155 is not found in either of the two models. Points from different data sets are offset slightly in multipole for clarity.}\label{fig:2}
\end{figure}
\begin{figure}
\centering
\includegraphics[width=8cm, height=5cm]{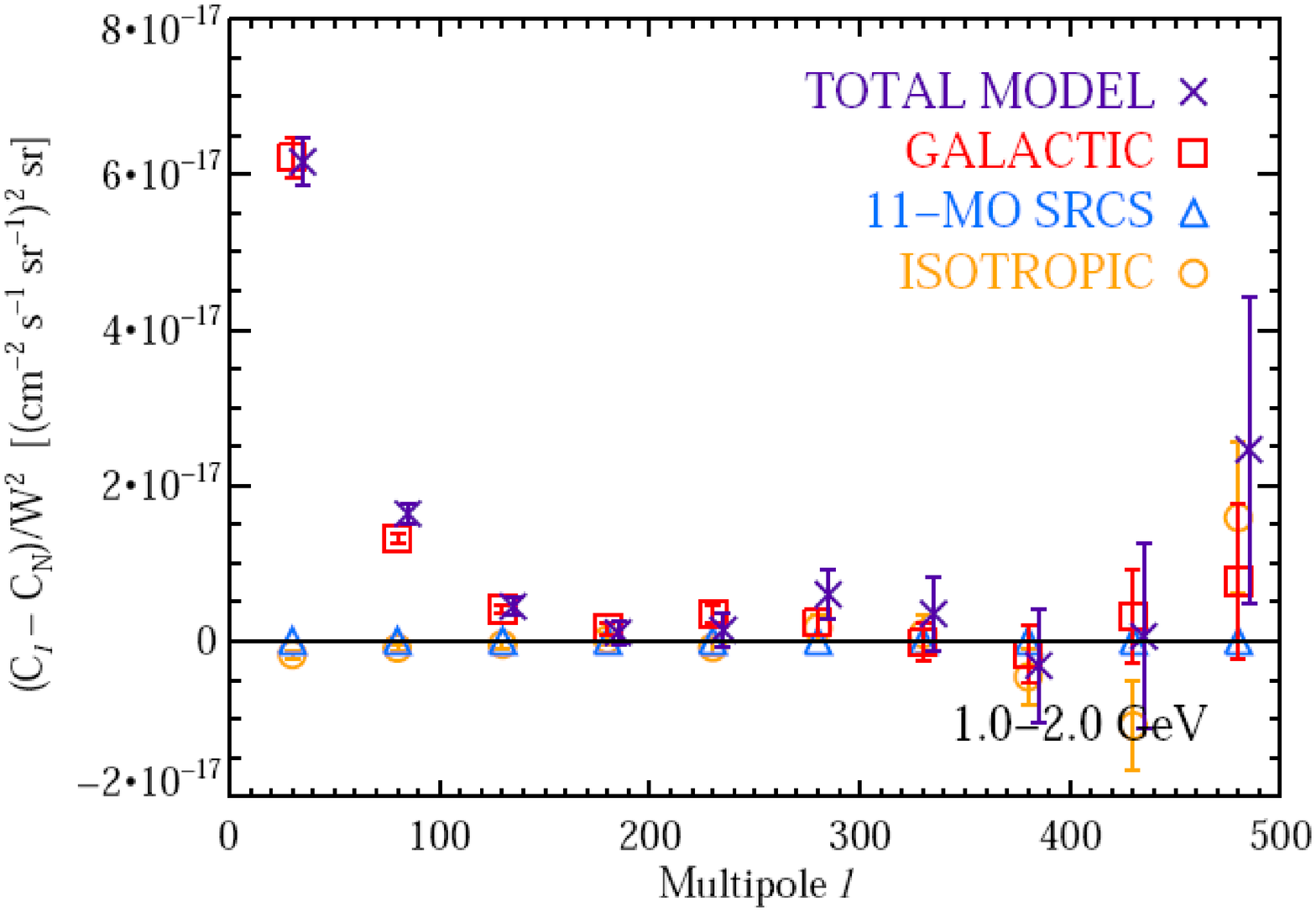}
\includegraphics[width=8cm, height=5cm]{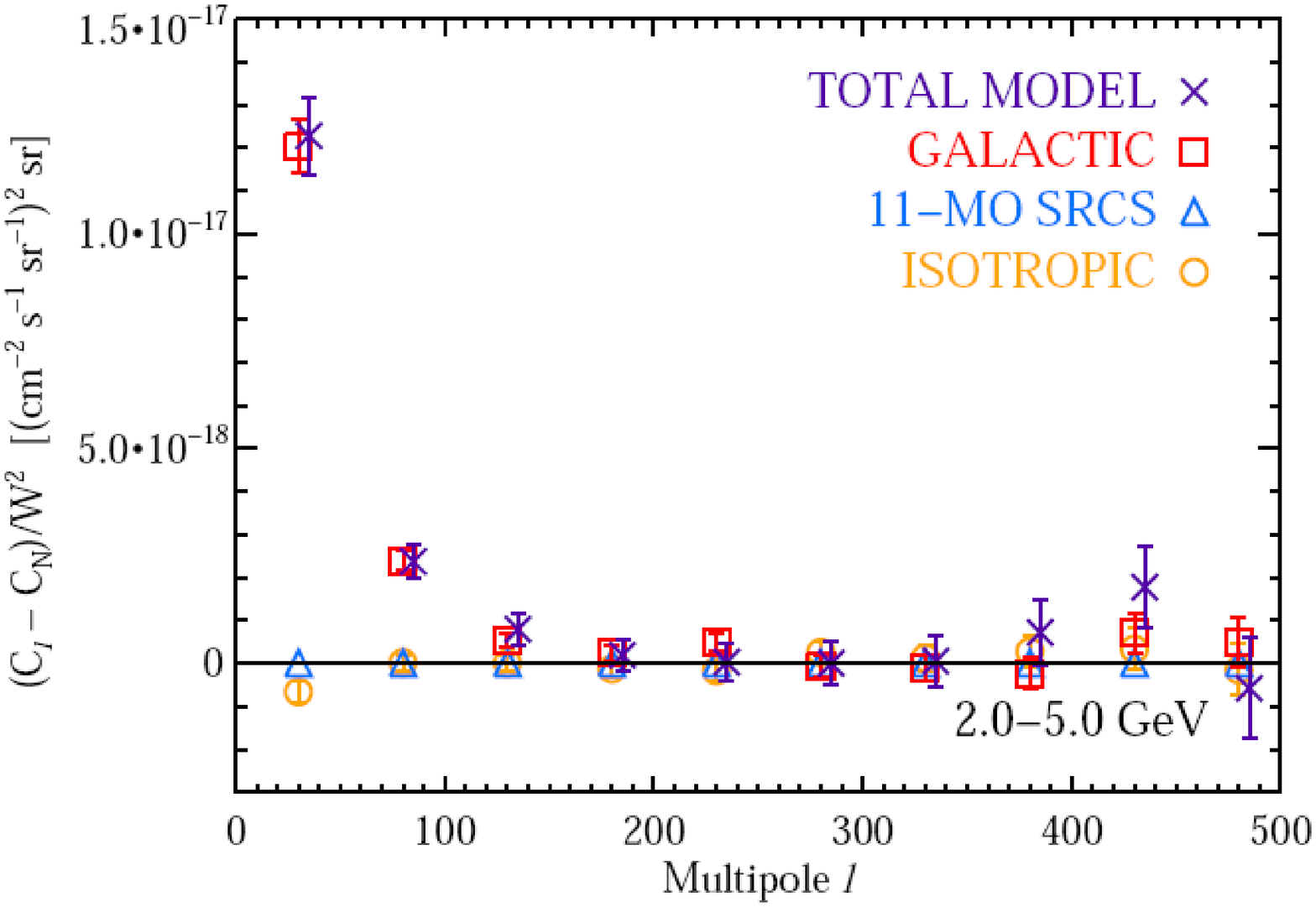}
\includegraphics[width=8cm, height=5cm]{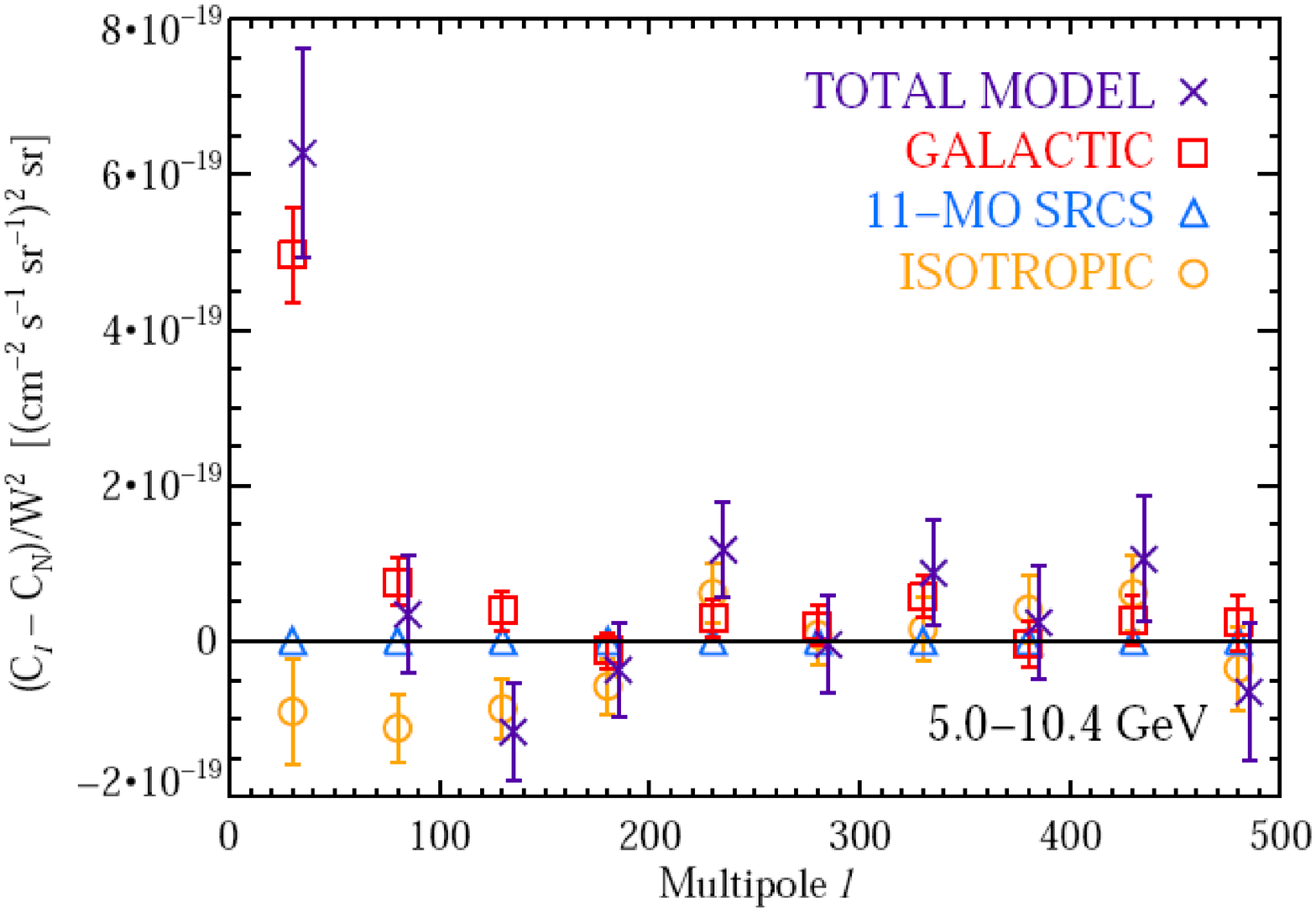}
\includegraphics[width=8cm, height=5cm]{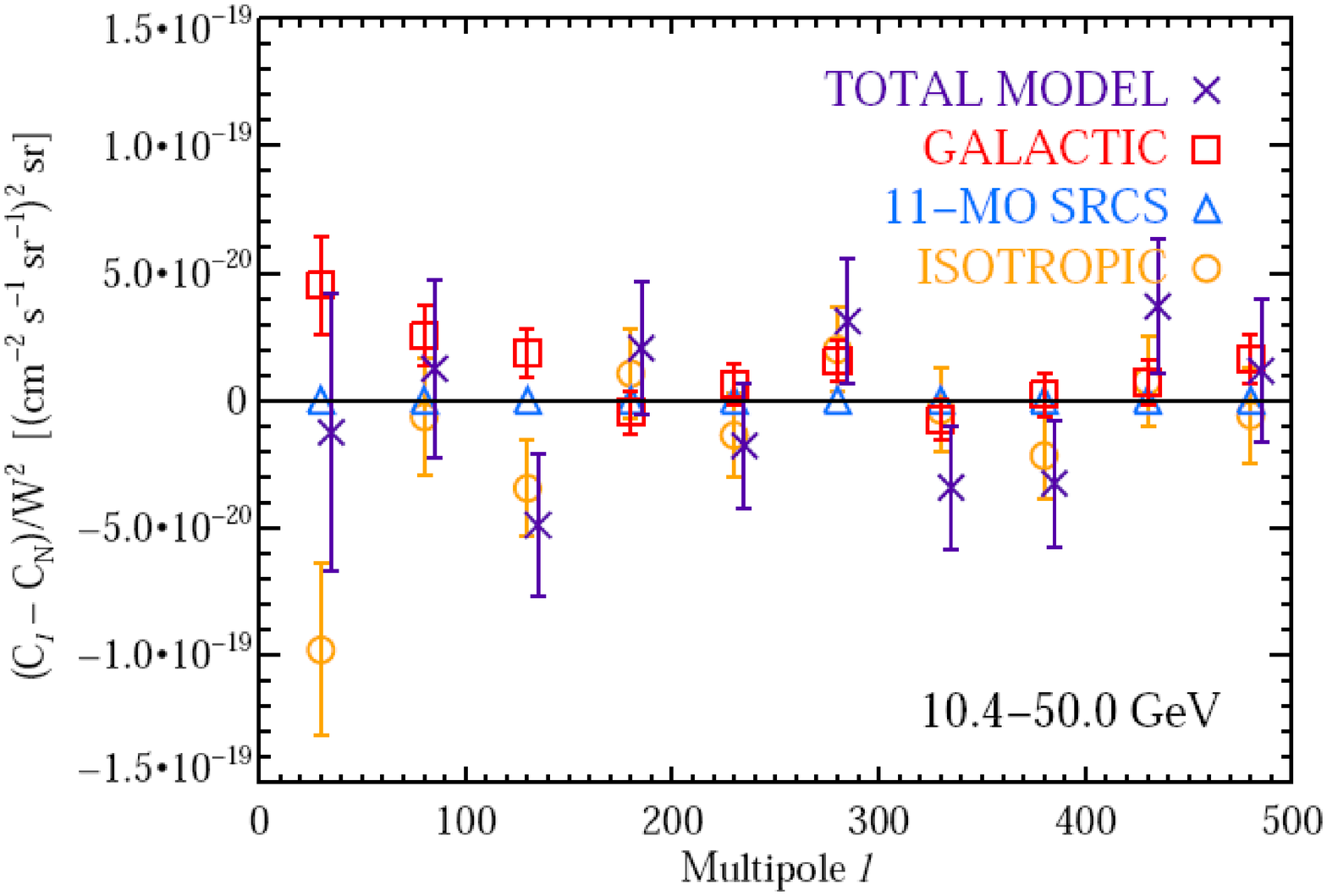}
\caption{PRELIMINARY. Intensity APS of the IGRB minus photon noise, in four energy bins. The APS of the single model components are reported. The large power in the models at $l<$ 155 is due to the GAL component. The isotropic component (ISO)  provides  APS compatible with zero as expected for  isotropic emission; likewise, the source component (CAT) provides no contribution because all sources were effectively masked. Total Model points are offset slightly in multipole for clarity.}\label{fig:3}
\end{figure}

\section{Simulated Models}

Detailed Monte Carlo simulations of Fermi-LAT all-sky observations were performed.
The purpose was  to have  a reference model to be  compared with the real data set. 
The status-of-art of the LAT modelling was used by means of the gtobssim tool. 
This code required as an input a detailed spatial and spectral model of the 
emission to be simulated.
Details of real LAT observations, such as  spacecraft pointing and live-time history, can be also included in the simulations. 
The gtobssim tool generates simulated photon events. 
The P6\_V3\_DIFFUSE IRFs and the actual spacecraft pointing and live-time history matching the observational time
 interval of the data were used to generate the simulated data sets. Two models of the gamma-ray sky were simulated.
Each model is the sum of three components: 
\begin{enumerate}
\item GAL - a model of the Galactic diffuse emission; 

\item CAT - the 1451 sources in the first Fermi-LAT source catalog (1FGL) \cite{latcal1}; 

\item ISO - an isotropic background.
\end{enumerate}

The same CAT and ISO components are included in both models.
These  differ only for the model for the GAL component in use. 
GAL describes both the spatial distribution and the energy spectrum of the Galactic diffuse emission. 
The GAL component for the reference sky model used in this analysis (hereafter, Model) is the recommended Galactic
 diffuse model for LAT data analysis, gll iem\_v02.fit \cite{galmod}, which has an angular resolution of 0.5$^{\circ}$. 
This model was used to obtain the 1FGL catalog; a detailed description can be found in \cite{galref}.

An  higher-resolution model (Hi-Res Model) was simulated for comparison, so that was  possible to test
impact of smaller details in the Galactic diffuse emission. This model (ring 21month v1.fit) is internal to the LAT collaboration,
 and was built using the same method as gll iem\_v02.fit, but differs primarily in the following ways: (i) this model was 
constructed using 21 months of Fermi -LAT observations, while gll iem\_v02.fit was based on 9 months of data;
(ii) the grid angular resolution for this model is 0.125$^{\circ}$ , in order fully exploit the angular resolution of the
 CO maps \cite{comap} used to build it; and (iii) additional large-scale structures, such as the Fermi bubbles \cite{bubbles}, are included
 in the model through the use of simple templates.

The single power law spectrum is assumed for all the  sources in CAT and 
the locations, average integral fluxes, and photon spectral indices are as reported in the 1FGL catalog.
 All 1451 sources were included in the simulation. ISO represents the sum of the Fermi-measured IGRB and an additional
 isotropic component presumably due to unrejected charged particles; for this component the spectrum template isotropic
 iem\_v02.txt was used. For both the Model and the Hi-Res Model , the sum of the three simulated components results in a
 description of the gamma-ray sky that closely approximates the angular-dependent flux and energy spectrum  of the all-sky
 emission measured by the Fermi -LAT. Although the simulated models may not accurately reproduce some large-scale structures, 
e.g., Loop I \cite{loop1} and the Fermi bubbles \cite{bubbles}, these features are not expected to induce anisotropies on the small angular scales on which
we focus in this work.

 The simulated models were processed and their angular power spectra calculated using the same analysis pipeline as the data.
In  Figure 2  the APS of the data and models are shown.
The contributions to APS of the individual components of the default model are shown  in Figure 3. 
At all energies the only component of the models contributing significantly to $l<155$ is the Galactic diffuse emission. 
The contribution from the isotropic component is negligible, since this component is isotropic
by construction and thus, after the photon noise is subtracted, it should only contribute to the monopole ($l$=0)
term.
 The source catalog component contributes zero power at all energies and multipoles because all these sources were
effectively masked, giving rise to a negligible residual effect.
We remark that in general the APS of distinct components are not linearly additive due to contributions from cross-correlations between the components.

\section{Results}

Significant  ($>$3$\sigma$ CL in each energy bin) angular power above the photon noise level (see Figure 1) is detected in the data at multipoles 155 $\le l \le$ 504 for energies below 10~GeV, and at lower significance in the 10-50~GeV energy bin. 
The sensitivity of the measurement at high energies is limited by the small statistics.
The measured angular power  is consistent with a constant value within each energy bin, which suggests that it originates from one or more unclustered populations of point sources. 
The fluctuation angular power detected in this analysis falls below the level predicted for some models of blazars, MSPs, and Galactic and extragalactic DM structures, and so the measured amplitude of the fluctuation angular power can limit the contribution to the total IGRB intensity of each source class.
The measured fluctuation angular power is almost independent of energy, and so  it might originate from a single dominant source class, although a mild energy dependence  cannot be excluded.

The energy dependence of the intensity angular power of the data is well-described by that arising from a single source class
 with a power-law energy spectrum with photon index $\Gamma$= −2.40$\pm$ 0.07. This value is compatible with the  mean intrinsic spectral index for blazars as determined from recent Fermi-LAT measurements.

The study of the currently available Fermi-LAT data has provided us with the first detection of small scale-angular power in the IGRB.
Future analyses on larger data sets could allow anisotropy searches to be extended to higher energies, and will likely be more sensitive to features in the anisotropy energy spectrum.

Detailed studies of population models  are required in order to:
(i)  identify the source classes which are actually contributing to the IGRB anisotropy and 
(ii)  provide upper-limits to the contribution of other candidate source populations.
A possible approach is the study of the APS induced by source population models and the 
following Monte Carlo simulation of the model emission, with the best possible characterization of the instrument response.

For an example,  in the  DM case a model of  all its distribution structures and sub-structure (Galactic and in the local Universe) 
which can provide contribution should be considered \cite{fornasa}, and
furthermore a given particle physics model should be assumed, or at least the DM particle mass and 
the annihilation cross-section for self-annihilating DM, or the mean life-time for pseudo-stable decaying DM.
Then the APS induced by these DM models can be obtained with the same procedure which was used for the simulated 
Model and Hi-Res Model of section 3.

\
\

{\bf Acknowledgements.}
The \textit{Fermi} LAT Collaboration acknowledges generous ongoing support
from a number of agencies and institutes that have supported both the
development and the operation of the LAT as well as scientific data analysis.
These include the National Aeronautics and Space Administration and the
Department of Energy in the United States, the Commissariat \`a l'Energie Atomique
and the Centre National de la Recherche Scientifique / Institut National de Physique
Nucl\'eaire et de Physique des Particules in France, the Agenzia Spaziale Italiana
and the Istituto Nazionale di Fisica Nucleare in Italy, the Ministry of Education,
Culture, Sports, Science and Technology (MEXT), High Energy Accelerator Research
Organization (KEK) and Japan Aerospace Exploration Agency (JAXA) in Japan, and
the K.~A.~Wallenberg Foundation, the Swedish Research Council and the
Swedish National Space Board in Sweden.

Additional support for science analysis during the operations phase is gratefully
acknowledged from the Istituto Nazionale di Astrofisica in Italy and the Centre National d'\'Etudes Spatiales in France.





\bibliographystyle{model3-num-names}
\bibliography{<your-bib-database>}



\end{document}